\documentclass{article}
\usepackage{spconf,amsmath,graphicx}
\usepackage[pdftitle={Noise-Robust DSP-Assisted Neural Pitch Estimation with Very Low Complexity},pdfauthor={Krishna Subramani, Jean-Marc Valin, Jan Buethe, Paris Smaragdis, Mike Goodwin}]{hyperref}
\usepackage{balance}
\usepackage{enumitem}
\usepackage{stmaryrd}
\ninept

\title{Noise-Robust DSP-Assisted Neural Pitch Estimation \\ with very low complexity}
\name{Krishna Subramani\sthanks{\ \ Work performed while at Amazon Web Services}\ $^{\flat}$, Jean-Marc Valin$^\natural$, Jan B{\"u}the$^\natural$, Paris Smaragdis$^{\flat\natural}$, Mike Goodwin$^\natural$
\address{$^\flat$ University of Illinois at Urbana-Champaign\quad$^\natural$ Amazon Web Services\\
\small{\texttt{ks51@illinois.edu, \{jmvalin, jbuethe, parsmara, mmg\}@amazon.com}}}}

\begin{document}
\maketitle
\begin{abstract}
Pitch estimation is an essential step of many speech processing algorithms, including speech coding, synthesis, and enhancement. 
Recently, pitch estimators based on deep neural networks (DNNs) have been outperforming well-established DSP-based techniques. Unfortunately, these new estimators can be impractical to deploy in real-time systems, both because of their relatively high complexity, and the fact that some require significant lookahead. We show that a hybrid estimator using a small deep neural network (DNN) with traditional DSP-based features can match or exceed the performance of pure DNN-based models, with a complexity and algorithmic delay comparable to traditional DSP-based algorithms. We further demonstrate that this hybrid approach can provide benefits for a neural vocoding task.
\end{abstract}
\section{Introduction}
\label{sec:intro}

Algorithms for estimating pitch based on DSP techniques have been around since the 1960s, and are still widely used in real-time systems to this day. These traditional DSP-based estimators include methods based on the cross-correlation~\cite{un1977pitch}, on the cepstrum~\cite{noll1964short}, and on the instantaneous frequency~\cite{mnasri2021novel}. While these methods are computationally efficient and generally produce acceptable results, many are subject to errors such as pitch period doubling. They can also be difficult to properly tune to minimize these errors while maximizing robustness to noise. 

Han et al.~\cite{han2014neural} were among the first to propose using neural networks to estimate the pitch in a supervised fashion. They propose using deep feedforward and recurrent neural networks to predict a log-scale quantized frequency value from spectral features -- effectively treating pitch estimation as a supervised classification problem. Kim et al. introduced CREPE~\cite{kim2018crepe}, a convolutional model for pitch estimation which was the first end-to-end model to learn pitch from the waveform. CREPE has demonstrated that it is remarkably robust to a wide variety of real world conditions and to noise. CREPE has been followed by further work on end-to-end pitch estimation, such as FCNF0~\cite{Ardaillon2019}, DeepF0~\cite{singh2021deepf0} and Penn~\cite{morrison2023cross}. In contrast, HarmoF0~\cite{wei2022harmof0} and~\cite{han2014neural} use frequency-domain input features, whereas~\cite{blok2021ife} uses the instantaneous frequency.

While machine learning based estimators are able to out-perform DSP-based techniques, they can be impractical to deploy in real-time systems. This is due to their high complexity, and the fact that some require significant lookahead (about 30~ms for CREPE).

We propose a hybrid approach that uses efficient DSP-derived features to make the pitch estimation task easier for a small DNN. Among potential features that can be used to estimate the pitch, we focus on the cross-correlation and the instantaneous frequency (Section~\ref{sec:features}). We propose DNN models (Section~\ref{sec:models}) that are designed to take advantage of those features -- either in isolation or in combination. We use the existing CREPE model as ground-truth for training our algorithm (Section~\ref{sec:data}). We show that the proposed approach can match or exceed the performance of pure DNN-based models, with a complexity and algorithmic delay comparable to traditional DSP-based algorithms (Section~\ref{sec:results}).

\section{Input Features}
\label{sec:features}
A DNN with sufficient capacity can be used to approximate almost any signal processing transformation. That is clearly demonstrated by CREPE~\cite{kim2018crepe}, which is able to estimate the pitch from a time-domain waveform. However, many transformation cannot be \textit{efficiently} implemented using DNN layers. For example, learning a discrete Fourier transform would require $O(N^2)$ operations rather than the $O(N\log N)$ complexity of the FFT. Similarly, since the cross-correlation requires multiplying the signal by itself, it cannot be implemented as a single layer and has to rely on the non-linear activation function to approximate the multiplicative effect. For these reasons, we propose starting from features that are already known to be useful for pitch estimation: the LPC residual cross-correlation~\cite{un1977pitch} and the instantaneous frequency~\cite{blok2021ife}. We make the assumption that these features still contain most of the relevant pitch information from the time-domain signal, while allowing us to use a much smaller DNN than would otherwise be needed.

Let $x[n]$ represent our signal. 
The short-time cross-correlation (Xcorr) for time-lag $\tau$ of a signal $x[n]$ with window length $N$ and hop size $H$ is defined as,
\begin{align}\label{eq:xcorr}
    R_{x}[m,\tau] = \sum_{n = 0}^{N - 1} x[mH + n]x[mH + n - \tau]\,.
\end{align}
We further normalize the above by the sum of the squared norm of the two sequences as explained in \cite{vos2013voice}. In practice, we find that computing the cross-correlation on the LPC prediction residual produces a more accurate pitch estimate, so that is what we use throughout this work. We only consider integer lags $\tau$, which means that the cross-correlation will have a higher relative accuracy for low pitch values (large periods) than for high pitch values where the integer lag becomes a limiting factor. 

The instantaneous frequency features are directly inspired from frequency reassignment (FR) \cite{flandrin:hal-00414583}. The idea behind FR is to refine frequency estimates by using the time-derivative of the phase. 
The main issue is that the reassigned frequencies are no longer constrained to be regular, thus preventing their usage with neural networks that implicitly assume the data to lie on a regular grid. What we propose instead is to work directly with the discrete time derivative of the STFT phase, which is simply a shifted version of the actual reassigned frequency \cite{flandrin:hal-00414583}. While this is not the exact expression for reassignment, we rely on our DNN to learn a more expressive map from these features for efficient pitch estimation.
To avoid the phase wrapping discontinuity between $\pi$ and $-\pi$, we use a normalized complex representation of the phase difference
\begin{align}
    \Delta_{x}[m,k] &= \frac{\delta_{x}[m,k]}{\left|\delta_{x}[m,k]\right|}\,, \label{eq:del_phi}\\
    \delta_{x}[m,k] &= F_{x}[m,k] \cdot F_{x}[m - 1,k]^{*}\,,
\end{align}
where $^*$ denotes the complex conjugation and $F_{x}[m,k]$ denotes the short-time Fourier transform (STFT) of $x[n]$ with window $w[n]$ of size $N$ and hop size $H$ such that
\begin{align}\label{eq:stft}
    F_{x}[m,k] = \sum_{n = 0}^{N - 1} w[n]x[mH + n]e^{\frac{-j2\pi kn}{N}}\,.
\end{align}

Our instantaneous-frequency (IF) features include the log-magnitude spectrum, as well as the real and imaginary components of $\Delta_{x}[m,k]$ in (\ref{eq:del_phi}). It is worth noting that through the inclusion of the log-magnitude spectrum, our IF features are theoretically sufficient to compute the cepstrum, which is also a useful pitch estimation feature. Unlike the Xcorr features, the IF features are most accurate for high pitch values where the harmonics are clearly separated in the STFT. For lower pitch values, extracting an accurate pitch out of the spectral features is more challenging. 

For both Xcorr and IF, we use $N = 320$ and $H = 160$. We choose to use a rectangular window due to its narrower main lobe and the fact that we do not need the high frequencies that would be affected by spectral leakage. The sampling rate for our experiments is $F_s = 16\ \textrm{kHz}$. We compute the cross-correlation over $\tau \in [0, 256]$, making the dimensionality of our Xcorr features 257. For the IF, we restrict ourselves to the first 30 discrete frequencies, thus making our IF dimensionality 90. 
We propose three different strategies: learning the pitch individually from both the Xcorr or IF features as well as combining the two to obtain a joint IF-Xcorr pitch estimator.



\section{Models}
\label{sec:models}

\begin{figure}[t]
    \centering
    \includegraphics[width=0.8\columnwidth]{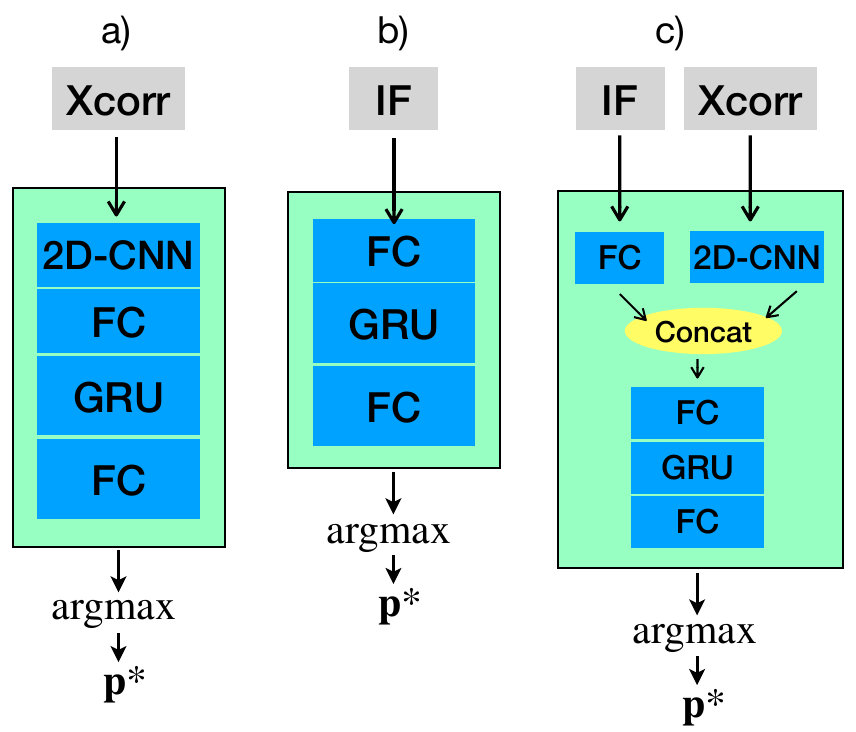}
    \caption{Network Architectures for pitch estimation from a) Cross-correlation (Xcorr), b) Instantaneous Frequency (IF) and c) Joint (IF + Xcorr) features. All networks output a distribution over the possible pitch values, the pitch estimate $\textrm{p}^{*}$ for a frame is the $\textrm{argmax}$ of the network output.}
    \label{fig:pitch_model}
\end{figure}

CREPE uses stacked 1D convolutional layers (CNN) on the input waveform to predict the pitch distribution. Since it is directly estimating pitch from the waveform, its first layers have to learn a relevant representation for estimating the pitch. Borrowing some knowledge from signal processing, we already know that the Xcorr and IF are robust features for pitch estimation. We use them as input features, and propose architectures that can best take advantage of these features. Traditional Xcorr-based pitch estimators attempt to find the peak in the noisy cross-correlation in a similar way that object detection works for images. For that reason, we propose using stacked 2D causal convolutional layers to do the Xcorr peak-finding. When it comes to the IF-based features, there is no ``local object'' to detect, but rather we need to consider all frequencies simultaneously. For that reason, we argue that fully-connected (FC) layers are a more appropriate way of processing the IF information.

Since the pitch of speech is usually smooth, it is common to use Viterbi-like post-processing to enforce some temporal consistency in the output~\cite{talkin1995robust}. However, instead of hardcoding the temporal dynamics and transition probabilities, we believe that recurrent networks -- more specifically gated recurrent units (GRU) -- can adequately model the temporal dynamics. Based on that, we propose three different network architectures depending on the input features (Fig.~\ref{fig:pitch_model}):
\begin{enumerate}[label=\alph*)]
    \item \textbf{Xcorr}: 3 causal 2D CNN layers of size 257 with $3\times3$ kernels and output 8 channels (except for the last layer). The CNN output goes to a 64-dimensional FC bottleneck, followed by a size-64 GRU. The GRU output goes to an FC layer of size 192. The total number of trainable parameters is 54689.
    \item \textbf{IF}: FC layer of input size 90 and output size 64, followed by a size-64 GRU and a FC layer of output size 192. The total number of trainable parameters is 47424.
    \item \textbf{Joint IF+Xcorr}: For the IF features, we use the same FC layer as b). For the Xcorr features, we use the same stack of CNN layers as a). The IF and Xcorr outputs are then concatenated and fed to a bottleneck FC layer with (64 + 257) inputs and 64 outputs, followed by a size-64 GRU and a FC layer of output size 192. The total number of trainable parameters is 68769.
\end{enumerate}

We use $\tanh(\cdot)$ as the activation in intermediate layers and softmax on the final output. We work with input batch sizes of 256, and training sequence length of 100 frames (or 1 second). We use the Adam optimizer with a learning-rate of $10^{-3}$. All networks are trained for 10 epochs on a single V100 GPU and minimize the weighted categorical cross-entropy loss. We weigh the categorical cross-entropy by the voicing to ignore estimates for the unvoiced frames.

\begin{figure}[h]
    \centering
    \includegraphics[scale=0.5]{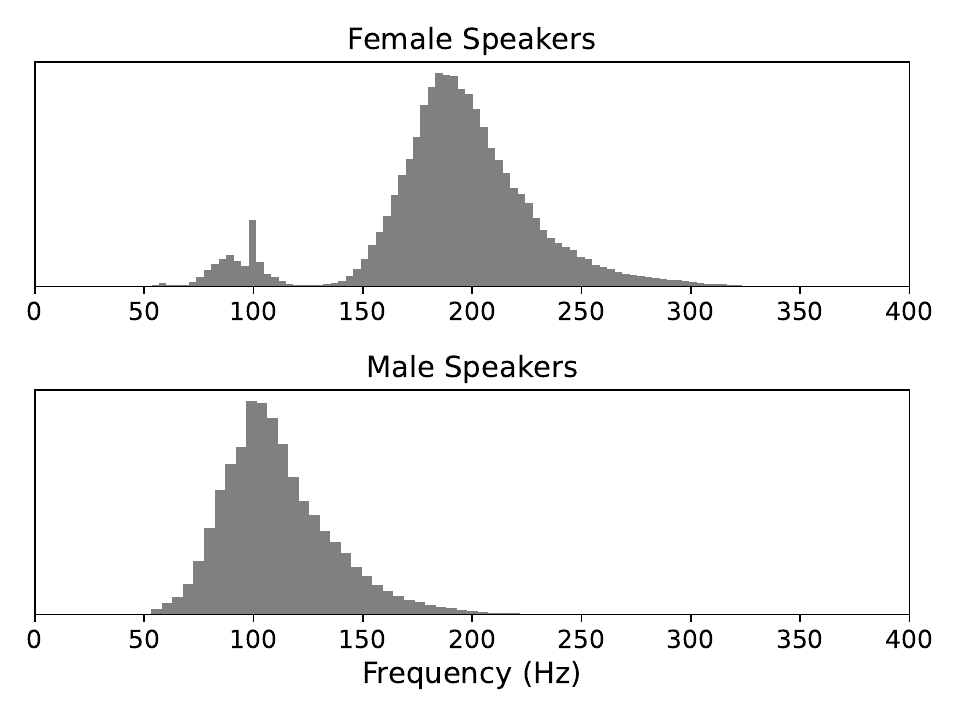}
    \caption{Histogram of PTDB reference pitch for female and male speakers. The bump in the female histogram below 125~Hz is due to period doubling errors.}
    \label{fig:ptdb_histogram}
\end{figure}

\section{Data and Evaluation}
\label{sec:data}
As with most machine learning-based approaches, we need labelled data to train our models, which is a challenge for speech. The standard dataset for labelled pitch for speech is the PTDB dataset~\cite{ptdb}, which provides the raw audio and the reference laryngograph pitch estimate. We observed however that the reference pitch estimates are not always reliable. We can see this by plotting the histogram of the reference pitches for male and female voices, as shown in Fig.~\ref{fig:ptdb_histogram}. We see a second peak for female speakers around 100~Hz, which we verified to be caused by period doubling on the reference laryngograph pitch estimation. Furthermore, we observe erroneously marked reference pitch estimates for unvoiced and silent sections. 
Thus, we avoid using the PTDB dataset to train our models. However, we use it for evaluation, but only after removing the female reference pitches below 125 Hz, and by performing energy thresholding to remove the reference pitches for unvoiced sections.

As a proxy metric to see how well our pitch estimator works, we use the raw cent accuracy (RCA), defined as the fraction of the frames where the predicted pitch lies within a 50-cent (half a semitone) interval of the reference ground truth pitch,
\begin{align}
    \textrm{RCA} = \frac{\sum_{\textrm{voiced}}\llbracket \,|\textrm{output} - \textrm{reference}| < 50 \,\rrbracket}{N_{\textrm{voiced}}},
\end{align}
where $\llbracket \,P\, \rrbracket$ is the Iverson bracket that returns 1 when $P$ is true and 0 otherwise, and $N_{\textrm{voiced}}$ is the total number of voiced frames. The output pitch and reference pitch in the expression above are both in cents. The cent scale is a logarithmic frequency scale, given by the transformation,
\begin{align}
    f_{\textrm{cent}} = 1200\cdot\log_{2}\left( \frac{f_{\textrm{Hz}}}{f_{0}}\right),
\end{align}
where $f_{0}$ is the reference frequency (62.5 Hz in our case, corresponding to a period of 256~samples at 16~kHz).

CREPE is not trained on PTDB, however it performs quite well on pitch estimation for PTDB (CREPE's RCA on PTDB is over 90\%). We thus propose using the CREPE estimated pitch as our ground-truth labels for training. We run CREPE, along with Viterbi post-processing on approximately 200 hours of several open speech datasets \cite{demirsahin-etal-2020-open, kjartansson-etal-2020-open, guevara-rukoz-etal-2020-crowdsourcing, he-etal-2020-open, kjartansson-etal-tts-sltu2018, oo-etal-2020-burmese, van-niekerk-etal-2017, gutkin-et-al-yoruba2020} and use the CREPE pitch (quantized to the nearest 20 cent interval) as our ground-truth pitch for training (we use CREPE's confidence output and threshold it as a proxy for voicing). 

To make our models more robust to noise, we augment our training data. For level augmentation, we randomly scale the input with gains lying in the [-60,10] dB range. For filtering, we generate random 2$^\textrm{nd}$ order IIR filters with coefficients lying in the [-3/8,\,3/8] range like in~\cite{valin2018rnnoise}. For additive noise, we use the Demand dataset~\cite{thiemann_joachim_2013_1227121} which contains multi-channel recordings of real-world noise. We use the first 4~minutes of noise for training and the last minute for evaluation. We keep 20\% of the training dataset unmodified, and augment the remaining 80\%. 

\section{Results}
\label{sec:results}

We test a total of 5~pitch estimation models: CREPE, the default estimator from the LPCNet vocoder, as well as the 3 proposed algorithms\footnote{An implementation of the proposed algorithm is available at \href{https://gitlab.xiph.org/xiph/opus/-/tree/icassp2024}{https://gitlab.xiph.org/xiph/opus/-/tree/icassp2024}} (Xcorr, IF, Joint). The LPCNet pitch estimator (LPE)~\cite{valin19_interspeech} is a variant of the RAPT algorithm~\cite{talkin1995robust}, which maximizes the normalized cross-correlation of the LPC residual, followed by a Viterbi-like causal smoothing. For CREPE we use the full (non-causal) Viterbi-decoded pitch output (which is also what the network was trained on). For evaluating the models in different Signal-to-noise (SNR) ratio conditions, we use the Demand dataset again (separate split from training to ensure that the network has not been trained on evaluation noise). We compute the RCA on the entire PTDB dataset with a 50 cent threshold. We also note here that none of our models (or CREPE) have been trained on the PTDB dataset. 

Given a frame of 10 ms, all proposed models estimates the pitch at the first sample given, making their algorithmic
delay equal to 10 ms. For LPE, the algorithmic delay is also 10 ms, whereas CREPE's (pre-Viterbi) algorithmic delay is about 32 ms.

\begin{figure}
    \centering
    \includegraphics[scale=0.55,trim=0cm 0cm 0cm 0.7cm]{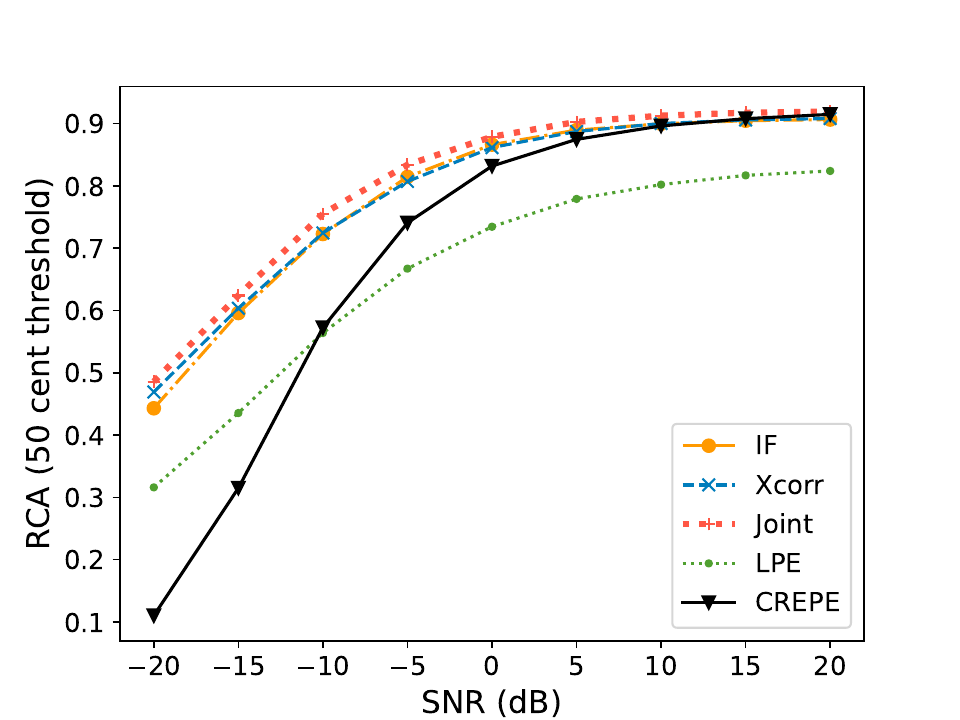}
    \caption{RCA for different SNR values on PTDB. All of our proposed models are significantly more robust to noise than CREPE, and also outperform the purely DSP-based LPE.}
    \label{fig:rca_snr}
\end{figure}

Fig.~\ref{fig:rca_snr} shows the variation of RCA with SNR for the 5 different pitch estimators, and Table~\ref{tab:rca_infty} shows the RCA computed for clean PTDB. We see that both the Xcorr and IF models perform similarly for different SNR values. Combining both clearly improves performance throughout and pushes the RCA on clean PTDB close to CREPE. All the neural pitch models perform better than LPE for all SNRs. The proposed models are also significantly more robust to noise than CREPE, especially at lower SNRs.

\begin{table}
    \centering
    \begin{tabular}{llllll}
    \hline
    Model & IF    & Xcorr & Joint & LPE & CREPE \\ \hline
    RCA (\%)   & 90.77 & 90.83 & 92.02 & 83.39  & 92.31 \\ \hline
    \end{tabular}
    \caption{RCA (higher is better) for models evaluated on clean PTDB (SNR $\rightarrow\, \infty$).}
    \label{tab:rca_infty}
\end{table}

\subsection{Complexity}
\label{sec:complexity}

Table~\ref{tab:complexity} compares the complexity of the proposed algorithms to that of the LPE and CREPE estimators. Even the most complex of the proposed models is about 5000x less complex than CREPE, despite performing similarly on clean conditions (Table~\ref{tab:rca_infty}) and better in noisy conditions (Fig.~\ref{fig:rca_snr}). As a comparison, even the CREPE ``tiny" model\footnote{\href{https://github.com/marl/crepe}{https://github.com/marl/crepe}} has a complexity of 7.4~GFLOPS. The proposed IF model has a complexity that is equivalent to that of the traditional LPE, despite being significantly more accurate in all tested conditions. Considering that the Xcorr model is almost as complex as the joint model, while having an accuracy similar to that of the IF model, we can conclude using the cross-correlation alone is not a good choice. In other words, instantaneous frequency features are always useful for pitch estimation -- regardless of whether a cross-correlation is also used.

\begin{table}[t]
    \centering
    \begin{tabular}{lccccc}
    \hline
    Model    & IF    & Xcorr & Joint & LPE & CREPE \\ \hline
    Features & 0.001 & 0.008 & 0.009 & 0.010  &  -    \\
    DNN      & 0.009 & 0.048 & 0.050 &   -    & 282   \\ \hline
    Total    & 0.010 & 0.056 & 0.059 & 0.010  & 282   \\ \hline
    \end{tabular}
    \caption{Complexity of the different pitch estimation algorithms, expressed in GFLOPS, where one multiply-add operation counts as two FLOPS. We separately evaluate the complexity of the feature computation and that of the DNN model.}
    \label{tab:complexity}
\end{table}

\subsection{Improved Neural Vocoding}
\label{sec:vocoding}
In many low-complexity vocoders, a key element is estimating the pitch of the input so that it is correctly replicated during synthesis.  If this is not done accurately, then we can observe pitch instability effects in the output speech, which negatively impacts the user experience. For that reason, we evaluate the proposed pitch estimators in the context of a neural vocoding task where the estimated pitch is used to resynthesize the input speech signal. We use the LPCNet~\cite{valin2019lpcnet} vocoder since it is designed to work with pitch features and uses LPE as its default estimator. The LPCNet models are trained following the training procedure specified in~\cite{subramani22_interspeech}, with the same public speech dataset, with the only difference being the pitch estimator. We measure the pitch mean absolute error (PMAE) of the synthesized speech, as described in Takahashi et al.~\cite{takahashi2023hierarchical} and Mustafa et al.~\cite{mustafa2023framewise}. The PMAE is defined as the mean $L_1$ norm between the pitch of the synthesized speech and that of the input speech, where the pitch is computed using the YAAPT algorithm~\cite{kasi2002yet}. Table~\ref{tab:pmae} shows the PMAE, along with the PESQ~\cite{P.862.2} values when evaluated on the clean PTDB dataset. All of the proposed estimators show a clear improvement over LPE -- both in terms of PMAE and PESQ, with the Xcorr and Joint models performing best overall. The accuracy of the proposed model does not significantly improve when increasing the complexity further.

\begin{table}
    \centering
    \begin{tabular}{lccccc}
    \hline
    Model & IF    & Xcorr & Joint & LPE & CREPE \\ \hline
    PMAE  & 3.301 & 3.214 & 3.232 & 3.500  & 3.289 \\
    PESQ  & 3.086 & 3.105 & 3.109 & 2.979  & 3.040 \\ \hline
    \end{tabular}
    \caption{PMAE (lower is better) and PESQ MOS-LQO (higher is better) for LPCNet synthesis on the PTDB dataset.}
    \label{tab:pmae}
\end{table}

\section{Conclusion}
This work demonstrates that accurate pitch estimation can be achieved using a small DNN combined with a sufficiently rich set of input features. In particular, the proposed system demonstrates the usefulness of instantaneous frequency features in determining the pitch. Results show that a combination of instantaneous frequency and cross-correlation features can achieve almost the same accuracy as an end-to-end DNN approach but with a significantly lower complexity, and that using just the instantaneous frequency alone makes it possible to match the complexity of traditional pitch estimators with a significantly higher accuracy, even in noisy conditions. We demonstrate through a speech synthesis task that the improved pitch estimation can provide benefits to real-life speech applications.


\balance
\bibliographystyle{IEEEbib}
\bibliography{refs}

\end{document}